\documentstyle[12pt]{article}
\catcode`\@=11

\@addtoreset{equation}{section}
\def\theequation{\arabic{section}.\arabic{equation}}

\def\@normalsize{\@setsize\normalsize{15pt}\xiipt\@xiipt
\abovedisplayskip 14pt plus3pt minus3pt%
\belowdisplayskip \abovedisplayskip
\abovedisplayshortskip  \z@ plus3pt%
\belowdisplayshortskip  7pt plus3.5pt minus0pt}

\def\small{\@setsize\small{13.6pt}\xipt\@xipt
\abovedisplayskip 13pt plus3pt minus3pt%
\belowdisplayskip \abovedisplayskip
\abovedisplayshortskip  \z@ plus3pt%
\belowdisplayshortskip  7pt plus3.5pt minus0pt
\def\@listi{\parsep 4.5pt plus 2pt minus 1pt
            \itemsep \parsep
            \topsep 9pt plus 3pt minus 3pt}}

\def\underline#1{\relax\ifmmode\@@underline#1\else
        $\@@underline{\hbox{#1}}$\relax\fi}
\@twosidetrue

\relax

\catcode`@=12

\catcode`\@=11

\def\section{\@startsection{section}{1}{\z@}{3.5ex plus 1ex minus
   .2ex}{2.3ex plus .2ex}{\large\bf}}

\def\thesection{\Roman{section}.}

\def\appendix{\setcounter{section}{0}
        \def\thesection{APPENDIX }
        \def\theequation{\Alph{section}.\arabic{equation}}}

\def\FERMIPUB{}

\def\ps@headings{\def\@oddfoot{}\def\@evenfoot{}
\def\@oddhead{\hbox{}\hfill
        \makebox[.5\textwidth]{\raggedright\ignorespaces --\thepage{}--
        \hfill {\rm FERMILAB--Pub--\FERMIPUB}}}
\def\@evenhead{\@oddhead}
\def\subsectionmark##1{\markboth{##1}{}}
}

\catcode`\@=12

\relax

\def\figcap{\section*{Figure Captions\markboth
        {FIGURECAPTIONS}{FIGURECAPTIONS}}\list
        {Fig. \arabic{enumi}:\hfill}{\settowidth\labelwidth{Fig. 999:}
        \leftmargin\labelwidth
        \advance\leftmargin\labelsep\usecounter{enumi}}}
 \relax
\def\tablecap{\section*{Table Captions\markboth
        {TABLECAPTIONS}{TABLECAPTIONS}}\list
        {Table \arabic{enumi}:\hfill}{\settowidth\labelwidth{Table 999:}
        \leftmargin\labelwidth
        \advance\leftmargin\labelsep\usecounter{enumi}}}
 \relax
\def\reflist{\section*{References\markboth
        {REFLIST}{REFLIST}}\list
        {[\arabic{enumi}]\hfill}{\settowidth\labelwidth{[999]}
        \leftmargin\labelwidth
        \advance\leftmargin\labelsep\usecounter{enumi}}}
 \relax

\catcode`\@=11

\def\FERMIPUB{}

\def\ps@headings{\def\@oddfoot{}\def\@evenfoot{}
\def\@oddhead{\hbox{}\hfill
        \makebox[.5\textwidth]{\raggedright\ignorespaces --\thepage{}--
        \hfill {\rm FERMILAB--Pub--\FERMIPUB}}}
\def\@evenhead{\@oddhead}
\def\subsectionmark##1{\markboth{##1}{}}
}

\ps@headings

\relax
\newskip\humongous \humongous=0pt plus 1000pt minus 1000pt

\newif\ifdtup

\def\beq{\begin{equation}}
\def\eeq{\end{equation}}

\def\beqn{\begin{eqnarray}}
\def\eeqn{\end{eqnarray}}
\relax

\def\G2{{\; \rm GeV/}c^2}
\def\G{\; \rm GeV}

\def\dotx{\dotx{\dot\overline{x}}}

\relax

\begin{document}
\hbadness=10000
\begin{titlepage}
\nopagebreak
\vspace*{1.5 cm}
\begin{flushright}

        {\normalsize
 KANAZAWA-93-07\\

   August,1993   }\\
\end{flushright}
\vfill
\begin{center}
{\large \bf Differential Calculi on h-deformed Bosonic and Fermionic Quantum
Planes}
\vfill
{\bf Tatsuo Kobayashi}

       Department of Physics, Kanazawa University, \\
       Kanazawa, 920-11, Japan \\
\vfill

\end{center}

\vfill
\nopagebreak
\begin{abstract}
We study differential calculus on h-deformed bosonic and fermionic quantum
space.
It is shown that the fermionic quantum space involves a parafermionic variable
as well as a classical fermionic one.
Further we construct the classical $su(2)$ algebra on the fermionic quantum
space and discuss a mapping between the classical $su(2)$ and the h-deformed
$su(2)$ algebras.

\end{abstract}

\vfill
\end{titlepage}
\pagestyle{plain}
\newpage
\voffset = -2.5 cm
\leftline{\large \bf 1. Introduction}
\vspace{0.8 cm}

Quantum groups and quantum algebras [1-4] have attracted much attention in
theoretical physics and mathematics, such as statistical models, integrable
models, conformal field theories, knot theory and so on [5-8].
Quantum space was introduced to represent quatum groups \cite {Manin}.
Differential calculus on the quantum space was also studied [10-15].
It is intriguing also from the viewpoint of non-commutative geometry.
The quantum differential calculus is closely related with q-oscillators,
which have been applied to various fields [16-18].
Further, we can obtain quantum groups $Sp_q(2n)$ and $SO_q(2n)$ using the
q-deformed phase space which is defined through the differential calculus
on the quantum space for $SL_q(n)$ \cite {Zumino}.
Furthermore, on the quantum space quantum deformed algebras have been
constructed, e.g., q-deformed Lorentz, Poincar{\'e}, conformal and
superconformal algebras [20-25].
These analyses imply that the quantum space is interesting as applications of
quantum groups and useful to show some aspects of quantum groups.

Recently other than the conventional q-deformation and its multi-parametric
extension, another deformation was discovered in [26-29] and is called
h-deformation.
Further, differential calculus on an h-deformed quantum bosonic space has been
 considered in \cite {Karimi,Agham}.
The purpose of this article is to investigate in detail bosonic and fermionic
differential calculus on h-deformed quantum planes, extending the above
analyses on the q-deformed differential calculus to the h-deformed case.
The differential calculus on the bosonic quantum space is studied and
a comment is obtained from the viewpoint of a constrainted system.
In addition, we discuss a differential calculus on an h-deformed fermionic
space.
Ref. \cite{Koba} shows that q-deformed fermionic coordinates and derivatives
represent $su_q(2)$ algebra, which is related with Drinfeld-Jimbo basis
by some mapping.
Following the approach, we investigate algebra which is constructed in terms
of the h-deformed fermionic elements.

This paper is organized as follows.
In section two we review on general non-commutative differential calculus
including the quantum space.
Then using a new solution of the Yang-Baxter equation, we derive h-deformed
bosonic differential calculus.
Further some comments on the h-deformed space are given.
In section three, similarly we derive h-deformed differential calculus on the
fermionic quantum space.
It is shown that we can represent the classical Lie algebra $su(2)$,
using their fermionic variables and derivatives.
Also we obtain a mapping between the classical algebra $su(2)$ and the
h-deformed algebra $su_h(2)$.
Further we discuss an h-deformed $SO(4)$ group.
The last section is devoted to conclusion and discussion.

\vspace{0.8 cm}
\leftline{\large \bf 2. Differential calculus on h-deformed bosonic space}
\vspace{0.8 cm}

A quantum space is a non-commutative space representing the corresponding
quantum group.
Ref. \cite{WZ} clarified general non-commutative differential calculus
including the quantum space and its analysis was extended to superspaces in
\cite{KU1}.
We set up commutation relations between coordinates $x^i$ and derivatives
$\partial_{x^i}$ as follows,
$$x^ix^j=B^{ij}_{\ \ k\ell}x^kx^\ell, \quad
\partial_{x^k} x^i=\delta^i_k+C^{ij}_{\ \ k\ell}x^\ell
\partial_{x^j}.
\eqno(2.1)$$
These matrices should satisfied the following relations:
$$B^{ij}_{\ \ pr}B^{rk}_{\ \ tn}B^{pt}_{\ \ \ell m}=
B^{jk}_{\ \ pr}B^{ip}_{\ \ \ell t}B^{tr}_{\ \ mn}, \quad
(\delta^i_k \delta^j_\ell-B^{ij}_{\ \ k\ell})(\delta^k_m\delta^\ell_n
+C^{k\ell}_{\ \ mn})=0.
\eqno(2.2)$$
The former equation is called Yang-Baxter equation.
We consider the case where $B^{ij}_{\ \ k\ell}$ is proportinal to
$C^{ij}_{\ \ k\ell}$.
In this case we can write commutation relations of derivatives by $B$-matrix
as follows,
$$\partial_{x^\ell}\partial_{x^k}=
B^{ij}_{\ \ k\ell}\partial_{x^j}\partial_{x^i}.
\eqno(2.3)$$

Recently a new solution of the Yang-Baxter equantion was discovered in
\cite{Ewen} as follows,
$$
 \widehat R =
\left(\begin{array}{cccc}
1 & -h' & h' & hh' \\
0 & 0 & 1 & h \\
0 & 1 &0 & -h  \\
0 & 0 & 0 & 1
\end{array}\right).
\eqno(2.4)$$
Also in Ref. \cite{Zak} and \cite{Dem}, the same type of the
$\widehat R$-matrix was discovered in the case where $h=h'$ and $h=h'=-1$,
respectively.
New type of a quantum group $SU_h(2)$ was studied in [26-28] and the
corresponding deformed algebra $su_h(2)$ has been constructed in \cite{Ohn}.
We identify $\widehat R$ with $B$ in order to obtain h-deformed differential
calculus.
The $\widehat R$-matrix has $\pm 1$ as eigenvalues.
This fact leads to a condition that we should identify $C$ with $\widehat R$,
too.
Namely we hereafter study the case where
$$ B^{ij}_{\ \ k\ell}=C^{ij}_{\ \ k\ell}=\widehat R^{ij}_{\ \ k\ell}.
\eqno(2.5)$$
We restrict ourselves to the case where $h=h'$.
It is easy to extend the following analysis to the case with two
independent parameters, $h$ and $h'$.
We substitute eq.(2.4) into (2.1) and (2.3), so that we obtain the following
commutation relations;
$$[x^1,x^2]=h(x^2)^2,\quad [\partial_{x^1},\partial_{x^2}]=h(\partial_{x^1})^2,
$$
$$[\partial_{x^2},x^1]=hx^1\partial_{x^1}+hx^2\partial_{x^2}
+h^2x^2\partial_{x^1},
\eqno(2.6)$$
$$[x^i\partial_{x^i},x^i]=1-hx^2\partial_{x^1}, \quad
[\partial_{x^1},x^2]=0.$$
The above algebra becomes classical in the limit, i.e.,
$h \rightarrow 0$.

We can introduce h-deformed quantum group $T^i_{\ j}$ such that the elements
transform the coordinates and the derivatives like $T^i_{\ j}x^j$.
Covariance of the commutation relations (2.6) under the transformation
requires the following commutation relation of $T^i_{\ j}$;
$$\widehat R^{ij}_{\ \ k\ell}T^k_{\ m}T^\ell_{\ n}=
T^i_{\ k}T^j_{\ \ell}\widehat R^{k\ell}_{\ \ mn}.
\eqno(2.7)$$

Next we consider conjugation consistent with (2.6).
Suppose that $h$ is pure imaginary, i.e., $\overline h=-h$.
Then we obtain the consistent conjugation as follows,
$$\overline {x^1}=x^1+hx^2, \quad \overline {x^2}=x^2,$$
$$ \overline {\partial_{x^1}}=-\partial_{x^1}, \quad
\overline {\partial_{x^2}}=-\partial_{x^2}-h\partial_{x^1}.
\eqno(2.8)$$
Through the conjugation we can define real coordinates and momenta as follows,
$$\widehat x^1=x^1+{h \over 2}x^2, \quad \widehat x^2 =x^2,$$
$$\widehat p_1=-i\partial_{x^1}, \quad
\widehat p_2=-i(\partial_{x^2}+{h \over 2}\partial_{x^1}).
\eqno(2.9)$$
They satisfy the following commutation relations;
$$[\widehat x^1,\widehat x^2]=h(\widehat x^2)^2, \quad
[\widehat p_1,\widehat p_2]=h(\widehat p_1)^2,$$
$$[\widehat p_j,\widehat x^j]=-i-h\widehat x^2\widehat p_1, \quad
[\widehat p_1,\widehat x^2]=0,
\eqno(2.10)$$
$$[\widehat p_2,\widehat x^1]=h(-i+\widehat x^1 \widehat p_1+
\widehat x^2 \widehat p_2-h\widehat x^2 \widehat p_1).$$
The similar phase space algebra as well as (2.6) has been obtained in
\cite{Karimi}.

These h-deformed coordinates and momenta can be represented in terms of
classical ones $\tilde x$ and $\tilde p$.
Suppose that $\widehat x^2=\tilde x^2$ and $\widehat p_1=\tilde p_1$, then we
have
$$\widehat x^1=\tilde x^1(1-ih\tilde p_1\tilde x^2)+ih(\tilde x^2)^2\tilde p_2,
 \quad
\widehat p_2=\tilde p_2(1-ih\tilde p_1\tilde x^2)+ih(\tilde p_1)^2\tilde x^2.
\eqno(2.11)$$

The deformed phase space algebra such as eq.(2.10) might remind us of Dirac
brackets of a constrainted system \cite{Dirac}.
 Actually we can find a \lq \lq constraint" which we can make identically
 equal to zero through the above commutation relations (2.10), as follows,
$$f=1-2ih\widehat p_1\widehat x^2.
\eqno(2.12)$$
The function $f$ commutes with $\widehat x^2$ and $\widehat p_1$ and satisfies
commutation relations with the other elements as follows,
$$[f,\widehat x^1]=i\widehat x^2 f,\quad [\widehat p_2,f]=i\widehat p_1f.
\eqno(2.13)$$
These relations seems to be somewhat different from the Dirac bracket, whose
right hand side vanishes completely.
But we can make $f$ equal to zero identically.
The above fact seems to show that $SU_h(2)$ is a `group' which transforms the
\lq \lq constraint" $f$ and its relations (2.13) covariantly.

\vspace{0.8 cm}
\leftline{\large \bf 3. Differential calculus on h-deformed fermionic space}
\vspace{0.8 cm}

In this section we discuss differential calculus on an h-deformed quantum
fermionic space.
Commutation relations of h-deformed coordinates $\theta^\alpha$ and
derivatives $\partial_\alpha$ are obtained by replacing $\widehat R$ in (2.1)
and (2.3) in terms of $-\widehat R$ as follows,
$$\theta^\alpha \theta^\beta=-\widehat R^{\alpha \beta}_{\ \ \mu \nu}
\theta^\mu \theta^\nu,
\quad \partial_\nu \partial_\mu =-\widehat R^{\alpha \beta}_{\ \ \mu \nu}
\partial_\beta \partial_\alpha,$$
$$\partial_\mu \theta^\alpha=\delta^\alpha_\mu
-\widehat R^{\alpha \beta}_{\ \ \mu \nu} \theta^\nu \partial_\beta.
\eqno(3.1)$$
These relations are written explicitly as follows,
$$(\theta^1)^2=h\theta^1\theta^2,\quad \{\theta^1,\theta^2\}=(\theta^2)^2=0,$$
$$(\partial_1)^2=\{\partial_1,\partial_2\}=0,\quad (\partial_2)^2=h\partial_1
\partial_2,$$
$$\{\partial_\alpha,\theta^\alpha \}=1+h\theta^2 \partial_1, \quad
\{\partial_1, \theta^2 \}=0,
\eqno(3.2)$$
$$\{\partial_2,\theta^1 \}=-h\theta^1 \partial_1-h\theta^2 \partial_2
-h^2\theta^2\partial_1.$$
Further eq.(3.2) leads to the following trilinear relations;
$$(\theta^1)^3=(\partial_2)^3=0.
\eqno(3.3)$$
The coordinate $\theta^1$ and the derivative $\partial_2$ are
parafermionic \cite{Para}, while its derivatives $\partial_1$ and its
coordinates $\theta^2$ are nilpotent.

As discussed in the previous section, we can introduce h-deformed quantum group
elements $T^\alpha_{\ \beta}$ which transform $\theta^\alpha$ into
$\theta'^\alpha=T^\alpha_{\ \beta}\theta^\beta$.
The elements satisfy the same commutation relation as (2.7).
This h-deformed quantum group is interesting also from the aspect that it
transforms fermionic and parafermionic variables, i.e., $\theta^1$ and
$\theta^2$.
Further, we can define a determinant of $T^\alpha_{\ \beta}$ using the
h-deformed fermionic space as follows,
$$\theta'^1 \theta'^2={\rm det} T \cdot \theta^1 \theta^2.
\eqno(3.4)$$
The definition leads to
$${\rm det}T=T^1_{\ 1}T^2_{\ 2}-T^1_{\ 2}T^2_{\ 1}+hT^1_{\ 1}T^2_{\ 1}.
\eqno(3.5)$$
Eq.(3.5) coinsides with the definition of the determinant in \cite{Dem}.

Ref.\cite{Koba} shows that q-deformed fermionic coordinates and derivatives
can represent q-deformed quantum algebra, e.g., $su_q(2)$.
Here we apply the similar analysis to the h-deformed case.
In the similar way to \cite{Koba}, we define the following generators;
$$L_+=\theta^2 \partial_1, \quad L_-=\theta^1 \partial_2.
\eqno(3.6)$$
Using their commutation relation, we introduce a Cartan generator as follows,
$$L_0=[L_+,L_-]=\theta^2\partial_2-\theta^1\partial_1.
\eqno(3.7)$$
In addition to (3.7), these generators satisfy the following relations;
$$[L_0,L_\pm]=\pm 2L_\pm.
\eqno(3.8)$$
Eqs. (3.7) and (3.8) are nothing but the classical $su(2)$ algebra up to the
normalization factor, although the coordinates and the derivatives are
deformed.
The generators act on the coordinates and the derivatives as follows,
$$[L_0,\theta^\alpha]=(-1)^\alpha \theta^\alpha, \quad
[L_0,\partial_\alpha]=-(-1)^\alpha \partial_\alpha,$$
$$[L_+,\theta^1]=\theta^2, \quad [L_+,\partial_2]=-\partial_1,$$
$$[L_+,\theta^2]=[L_+,\partial_1]=0,$$
$$[L_-,\theta^1]=-h\theta^1 L_0-h^1\theta^2 L_+,
\eqno(3.9)$$
$$[L_-,\theta^2]=\theta^1 +h\theta^2 L_0,$$
$$[L_-,\partial_1]=-\partial_2+h\partial_1 L_0+h\partial_1 L_0,$$
$$[L_-,\partial_2]=-h\partial_2 L_0-3h^2\partial_1+3h^2\partial_2 L_+.$$
The actions of $L_+$ and $L_0$ are never deformed, while those of $L_-$ are
deformed.

In Ref.\cite{Ohn} a h-deformed $su(2)$ algebra has been constructed as follows,
$$[H,X]={2 {\rm sinh} (hX) \over h},$$
$$[H,Y]=-\{Y,{\rm cosh} (hX) \},
\eqno(3.10)$$
$$[X,Y]=H.$$
The h-deformed algebra $su_h(2)$ can be related with the classical algebra
$su(2)$ through the following mapping;
$$X={{\rm log}\tilde L_+ \over h},$$
$$H=(\tilde L_+ -(\tilde L_+)^{-1})\tilde L_0, \eqno(3.11)$$
$$Y={h \over 2}(\tilde L_+ -(\tilde L_+)^{-1})(2\tilde L_+\tilde L_- -h
\tilde L_0),$$
where $\tilde L_0 =L_0/2$ and $\tilde L_\pm =L_\pm /\sqrt 2$.

In Ref.\cite{Zumino} Zumino derived quantum groups $Sp_q(2n)$ and $SO_q(2n)$
from
q-deformed bosonic and fermionic phase space algebra for $SL_q(n)$.
That was extended to the supersymmetric case in \cite{KU1}.
In the similar way, here we discuss h-deformation of $SO(4)$.
At first, we define h-deformed gamma matrices $\gamma^\alpha$ as
$$\gamma^\alpha = \partial_\alpha, \quad \gamma^{5-\alpha}=\theta^\alpha.
\eqno(3.12)$$
They satisfy the following h-deformed Clifford algebra;
$$ \gamma^\alpha \gamma^\beta +\tilde B^{\alpha \beta}_{\ \ \mu \nu}
\gamma^\mu \gamma^\nu=\eta^{\alpha \beta},
\eqno(3.13)$$
where $\eta^{\alpha \beta}$ is an SO(4) metric.
The matrix $\tilde B$ is composed of $\widehat R$-matrix (2.4) as follows,
$$\tilde B^{\alpha'\beta'}_{\ \ \mu' \nu'}=
-\widehat R^{\alpha \beta}_{\ \ \mu \nu}, \quad
\tilde B^{\alpha \beta}_{\ \ \mu \nu}=
-\widehat R^{\nu \mu}_{\ \ \beta \alpha},$$
$$\tilde B^{\alpha \beta'}_{\ \ \mu' \nu}=
-\widehat R^{\beta \nu}_{\ \ \alpha \mu }, \quad
\tilde B^{\alpha' \beta}_{\ \ \mu \nu'}=
-(\widehat R^{-1})^{\alpha \mu}_{\ \ \beta \nu},
\eqno(3.14)$$
where $\alpha,\beta,\mu,\nu=1,2$ and $\alpha'=5-\alpha$.
Ref.\cite{KU1} shows that $\tilde B$ constructed through the above procedure
satisfy the Yang-Baxter equation, if $\widehat R$ is the solution of the
Yang-Baxter equation.
We could introduce h-deformed $SO(4)$ group which transform the relation
(3.13) covariantly.
Instead of deriving explicitly $SO_h(4)$, we here introduce h-deformed
$SO(4)$ quantum space $X^i$ $(i=1\sim 4)$ whose commutation relations are
written as $X^iX^j=\tilde B^{ij}_{\ \ k \ell}X^kX^\ell$.
We have explicitly
$$[X^1,X^2]=h(X^1)^2, \quad [X^1,X^3]=0,$$
$$[X^1,X^4]=[X^2,X^3]=-hX^3X^1,
\eqno(3.15)$$
$$[X^2,X^4]=hX^4X^1+hX^3X^2+h^2X^3X^1, \quad
[X^3,X^4]=-h(X^3)^2.$$
The algebra has a center element $C\equiv X^4X^1+X^3X^2$.
Differential calculus on the above h-deformed space $X^i$ could be similarly
obtained.
Their commutation relations are covariant under the $SO_h(4)$ transformation,
as said the above.
Elements of $SO_h(4)$ satisfy the same relation as (2.7) except
$\widehat R$-matrix replacing $\tilde B$-matrix.

\vspace{0.8 cm}
\leftline{\large \bf 4. Conclusion}
\vspace{0.8 cm}

We have studied here the differential calculi on the h-deformed bosonic and
fermionic spaces.
We have constructed the classical $su(2)$ algebra on the h-deformed fermionic
space.
The algebra is related with the h-deformed algebra $su_h(2)$.
It is shown that $SU_h(2)$ is a 'group' which transforms fermionic and
parafermionic variables into each other.
This fact is very interesting in applications to the parastatistics.
Also $SO_h(4)$ was discussed.
Supersymmetric extension is also intriguing.
It is easy to introduce h-deformed oscillators in the similar way to the above
 h-deformed differential calculus.
For example, we can define a sort of deformed oscillators by identifying
the h-deformed coordinates and derivatives with deformed anihilation and
creation operators, respectively.
Their applications to various fields are very interesting.

\vspace{0.8 cm}
\leftline{\large \bf Acknowledgement}
\vspace{0.8 cm}

The author would like to thank T.~Uematsu for numerous valuable discussions
and reading the manuscript.
He also thanks to P.~P.~Kulish, R.~Sasaki and H.~Terao for helpful discussions.


\newpage
\begin{thebibliography}{99}


\bibitem{Drinfeld}
V.~G.~Drinfeld, Sov.~Math.~Dokl.~{\bf 32}(1985)254.

\bibitem{Jimbo}
M.~Jimbo, ~Lett.~Math.~Phys.~{\bf 10}(1985)63.

\bibitem{Faddeev-Reshetikhin-Takhtajan}
L.~D.~Faddeev, N.~Yu.~Reshetikhin and L.~A.~Takhtajan, Algebra and
Analysis, {\bf 1}(1987)178.

\bibitem{KR}
P.~P.~Kulish and N.~Yu.~Reshetikhin,
J.~Sov.~Math. {\bf 23}(1983)2435.

\bibitem{Doebner-Hennig}
H.~-D.~Doebner and J.~-D.~Hennig eds., {\it Quantum Groups},  Lecture Notes
in Physics {\bf 370}(Springer Verlag, 1990).

\bibitem{CFZ}
T.~Curtright, D.~Fairlie and C.~Zachos eds., {\it Quantum Groups},
Proceedings of the Argonne Workshop (World Scientific, 1990).

\bibitem{Tak}
L.~A.~Takhtajan, Adv.~Studies~Pure~Math.~{\bf 19}(1989)435; Lecture Notes in
Physics {\bf 370}(1989)3.

\bibitem{K}
P.~P.~Kulish ed., {\it Quantum Groups}, Lecture Notes in Mathematics
{\bf 1510}(Springer Verlag, 1992).

\bibitem{Manin}
Yu.~I.~Manin, Commun.~Math.~Phys.~{\bf 123}(1989)163.

\bibitem{Woro}
S.~L.~Woronowicz, Commun.~Math.~Phys.~{\bf 111}(1987)613; ~{\bf 122}(1989)125.
\hfill\break
Publ.~RIMS, Kyoto Univ. {\bf 23}(1987)117.

\bibitem{WZ}
J.~Wess and B.~Zumino, Nucl.~Phys.~B~(Proc.~Suppl.)~{\bf 18B}(1990)302.

\bibitem{Schirrmacher-Wess-Zumino}
A.~Schirrmacher, J.~Wess and B.~Zumino, Z.~Phys.~C {\bf 49}(1991)317.

\bibitem{Schirrmacher1}
A.~Schirrmacher, Z.~Phys.~C {\bf 50}(1991)321.

\bibitem{Carow-Watamura-Schlieker-Watamura}
U.~Carow-Watamura, M.~Schlieker and S.~Watamura, Z.~Phys.~C {\bf 49}(1991)439.

\bibitem{KU1}
T.~Kobayashi and T.~Uematsu, Z.~Phys.~C {\bf 56}(1992)
193.

\bibitem{Mac}
A.~J.~Macfarlane, J.~Phys.~A {\bf 22}(1989)4581.

\bibitem{Bied}
L.~C.~Biedenharn, J.~Phys.~A {\bf 22}(1989)L783.

\bibitem{ChKu}
M.~Chaiahian and P.~P~Kulish, Phys.~Lett. {\bf B234}(1990)72.

\bibitem{Zumino}
B.~Zumino, Mod.~Phys.~Lett. {\bf A6}(1991)1225.

\bibitem{SWZ}
W.~B.~Schmidke, J.~Wess and B.~Zumino, Z.~Phys.~C {\bf 52}(1991)471.

\bibitem{OSWZ2}
O.~Ogievetsky, W.~B.~Schmidke, J.~Wess and B.~Zumino, Lett.~Math.~Phys.
{\bf 23}(1991)233.

\bibitem{OSWZ}
O.~Ogievetsky, W.~B.~Schmidke, J.~Wess and B.~Zumino,
Commun. Math. Phys.~{\bf 150}(1992)495.

\bibitem{KU2}
T.~Kobayashi and T.~Uematsu, Z.~Phys.~C {\bf 58}(1993)559.

\bibitem{KU3}
T.~Kobayashi and T.~Uematsu, Phys.~Lett. {\bf B306}(1993)27.

\bibitem{Koba}
T.~Kobayashi, \lq \lq Quantum Deformed $su(m|n)$ Algebra and Superonformal
Algebra on Quantum Superspace" Preprint KUCP-59; to be published in Z.~Phys.~C.

\bibitem{Zak}
S.~Zakrzewski, Lett.~Math.~Phys. {\bf 22}(1991)287.

\bibitem{Ewen}
H.~Ewen, O.~Ogievetsky and J.~Wess, Lett.~Math.~Phys. {\bf 22}(1991)297.

\bibitem{Dem}
E.~E.~Demidov, Yu.~I.~Manin, E.~E.~Mukhin and D.~V.~Zhdanovich,
Prog.~Theor.~Phys.~Suppl. {\bf 102}(1990)203.

\bibitem{Ohn}
Ch.~Ohn, Lett.~Math.~Phys. {\bf 25}(1992)85.

\bibitem{Karimi}
V.~Karimipour, \lq \lq The Quantum de Rham Complexes Associatd with $SL_h(2)$"
Preprint.

\bibitem{Agham}
A.~Aghamohammadi, \lq \lq The 2-Parametric Extension of $h$ deformation of
$GL(2)$, and The Differential Calculus on Its Quantum Plane" Preprint.


\bibitem{Dirac}
P.~Dirac, {\it Lectures on Quantum Mechanics} (Yeshiva University, NY, 1964).

\bibitem{Para}
Y.~Ohnuki and S.~Kamefuchi, {\it Quantum Field Theory and Parastatistics}
(Springer Verlang, Berlin, 1982).

\end{thebibliography}
\end{document}